\documentclass[pra,aps,twocolumn,showpacs,epsfig]{revtex4}
\newcommand{\beq}{\begin{equation}}
\newcommand{\eeq}{\end{equation}}
\newcommand{\beqa}{\begin{eqnarray}}
\newcommand{\eeqa}{\end{eqnarray}}
\newcommand{\ba}{\begin{array}}
\newcommand{\ea}{\end{array}}
\usepackage{color}
\usepackage[dvips]{epsfig}

\begin{document}

\title{Quantum-tunneling dynamics of a spin-polarized Fermi gas in a
double-well potential}
\author{L. Salasnich$^{1}$, G. Mazzarella$^{1}$, M. Salerno$^{2}$,
and F. Toigo$^{1}$}
\affiliation{$^{1}$Dipartimento di Fisica ``Galileo Galilei'' and CNISM,
Universit\`a di Padova, Via Marzolo 8, 35122 Padua, Italy\\
$^{2}$Dipartimento di Fisica ``E.R. Caianiello'', CNISM  and INFN -
Gruppo Collegato di Salerno,
Universit\`a di Salerno, Via Ponte don Melillo, 84084 Fisciano (SA), Italy}

\begin{abstract}
We study the exact dynamics of a one-dimensional
spin-polarized gas of fermions in a double-well potential at zero and finite
temperature. Despite
the system is made of non-interacting fermions, its dynamics can be quite
complex, showing strongly
aperiodic spatio-temporal patterns during the tunneling.
The extension of these results to the case
of mixtures of spin-polarized fermions in interaction with self-trapped
Bose-Einstein condensates (BECs)
at zero temperature is considered as well. In this case we show that the
fermionic dynamics
remains qualitatively similar to the one observed in absence of BEC but
with the Rabi frequencies of fermionic
excited states explicitly depending on the number of bosons and on the
boson-fermion interaction strength. From this, 
the possibility to control quantum fermionic dynamics by means of  Feshbach
resonances is suggested.
\end{abstract}


\maketitle

\section{Introduction}

The control and manipulation of confined ultracold atomic gases make 
possible the study of the dynamics 
of many-body quantum states either for bosons \cite{folling,spielman,salger,
trotzky,wurtz} and for fermions \cite{shin,inguscio}, as well as for their
mixtures. Ultracold atoms
give the opportunity to investigate the most important quantum effects with a
very high level of control and precision \cite{pitaevskii,dalfovo,pethick,
leggett}. Among these effects, the quantum
tunneling is a topic of wide interest. In this sense, one of the most important
examples is provided by Bose gases confined in double-well shaped potentials.
In these systems the tunneling of bosons through the central barrier,
which separates the two potential wells, is responsible for the atomic
counterpart of the
Josephson effect  (see, for instance, Refs.
\cite{leggett,smerzi,albiez,mazzarella}) observed in superconductive
junctions \cite{book-barone}.
The macroscopic quantum tunneling was studied also with two weakly-linked
superfluids made of interacting fermionic atoms
for which it is possible to obtain atomic Josephson junction equations
\cite{salasnich1}.

The optical lattices and double-well traps are paradigmatic external
potentials to
understand the tunneling in atomic systems characterized by a small
number of particles \cite{folling,trotzky,esteve}.
The dynamics of small systems of spinless bosons \cite{zollner}, of finite
Fermi-Hubbard and Bose-Hubbard systems
\cite{ziegler}, and of spin $1/2$ fermions - especially within the quantum
information
community \cite{lewenstein,cirac} -  were widely studied.

In this work we consider a one-dimensional (1D) dilute and ultracold gas of
spin-polarized fermionic atoms
in a double-well potential in the absence and in the
presence of a Bose-Einstein condensate (BEC) which is intrinsically
localized in one of the two potential wells. In both cases we have that 
in correspondence to certain values of the
parameters of the double-well potential, its eigenspectrum is organized in
doublets spaced between themselves and made of
quasi-degenerate states. Each doublet is characterized by its own Rabi
frequency, proportional to the gap between the two
corresponding states. It turns out that even in absence
of BEC and of  interactions between the fermions,
this kind of system exhibits an interesting behavior.

In the first part of the paper we focus on a spin-polarized
quantum gas in absence of BEC. We suppose that at the initial
time the system is prepared in such a way that
all the fermions are localized at a given side of the barrier
and that a number of fermionic states can be excited by means of
external resonant fields (e.g. resonant with the Rabi frequencies of the
doublets). The dynamics sustained by the above
Hamiltonian allows to study the transfer of matter which takes place through
the barrier by 
tunneling. This effect was studied in Ref. \cite{zollner} for few bosons in
a 1D double-well at zero-temperature.
Indeed, the non-interacting pure fermionic system can be viewed as a
realization in the infinitely strong repulsive interaction limit
of 1D hard-core bosonic systems \cite{grd}, where the Pauli exclusion
principle mimics the hard-core interaction.
By following Ref. \cite{zollner}, we fix the number of fermions and analyze
the spatial variations of the density
profile of the fermionic cloud at different times, and the total fermionic
fractional imbalance $Z_F$ between the two wells as a function of  time $t$.
We carry out this analysis both with one fermion and
with more fermions by pointing out the striking differences between the
two cases. In the former case the temporal behavior of $Z_F(t)$ is fully
periodic and its period (Rabi period) is given by the inverse Rabi frequency
of the lowest doublet.
In the latter case the periodicity occurs on much longer
times, equal to the minimum common multiple of  the Rabi periods
of the populated doublets.  $Z_F(t)$  may even be completely aperiodic if some
of the Rabi periods of the populated doublets are incommensurate.
In any case,  the shape of the density profile generally exhibits spatio
temporal patterns which do not replicate the initial situation for rather
long times.
In Refs. \cite{zollner,ziegler}, the atomic systems therein considered are
dealt with at zero-temperature. Here we perform our analysis also at nonzero
temperature. 
By keeping fixed the number of fermions, we study the influence of the
temperature on the total fractional imbalance: 
there is a spatial broadenings of the density profile that
become more evident by increasing the temperature. 

In the second part of the paper we discuss the case of a spin-polarized
quantum Fermi
gas interacting with a quasi-stationary BEC at zero temperature.
In particular, we show that for a self-trapped BEC (e.g. a BEC intrinsically
localized by the nonlinearity in one
of the two wells of the potential) and for small boson-fermion interactions,
the fermionic imbalance dynamics
remains qualitatively similar to the one of the pure fermionic case with the
only difference that the spacings of the fermionic levels
(and therefore the Rabi frequencies) become dependent on both the number of
bosons and the boson-fermion interaction.
This suggests the possibility to control the fermionic quantum dynamics by
changing, for example, the inter-species scattering lengths
by using Feshbach resonances, a fact which could be of interest for
applications to quantum computing.

\section{The system}

We consider a confined dilute and ultracold spin-polarized gas of $N_F$
fermionic atoms of mass $m$ in a double-well potential. For dilute and
ultracold atoms the only active channel of the inter-atomic
potential is the s-wave scattering. For spin-polarized fermions
the s-wave channel is inhibited by statistics and consequently the
system is practically an ideal Fermi gas.
The trapping potential of the system is given by
\beq
V({\bf r}) = U(x) + {1\over 2} m \omega_{\bot}^2 (y^2 + z^2) \; ,
\eeq
We suppose that the
system is one-dimensional (1D), due to a strong radial transverse
harmonic confinement \cite{flavio1,flavio2}, with a symmetric double-well
trapping potential in the longitudinal axial direction, that we
will denote by $U(x)$. We are thus assuming that the transverse energy
$\hbar \omega_{\bot}$ of the confinement
is much larger than the characteristic energy
of fermions in the axial direction \cite{flavio1,flavio2}.

On the basis of the dimensional reduction,
the 1D Hamiltonian of the dilute spin-polarized Fermi gas confined by
the potential $U(x)$ is
\beq {\hat H} = \int {\hat
\psi}^{\dagger}(x) \left[ -{\hbar^2\over 2m} {\partial^2\over
\partial x^2} + U(x) \right] {\hat \psi}(x) \ dx \; , \label{ham}
\eeq
where ${\hat \psi}(x)$ and $ {\hat \psi}^{\dagger}(x) $ are
the usual field operators annihilating or creating a fermion at position
$x$, therefore obeying anticommutation rules. The single-particle
stationary Schr\"odinger equation associated to the Hamiltonian
(\ref{ham}) can be written as
\beq
\left[ -{\hbar^2\over 2m}
{\partial^2\over \partial x^2} +U(x) \right] \phi_{\alpha,j}(x) =
\epsilon_{\alpha,j} \phi_{\alpha,j}(x) \label{eigen} \;,
\eeq
where $\phi_{\alpha,j}(x)$ are the complete set of orthonormal
eigenfunctions and $\epsilon_{\alpha,j}$ the corresponding
eigenenergies. Here $\alpha=1,2,3,...$ gives the ordering number
of the doublets of quasi-degenerate states, and $j=S,A$ gives the
symmetry of the eigenfunctions ($S$ means symmetric and $A$ means
anti-symmetric). It follows that the eigenvalues are ordered in
the following way: $\epsilon_{1,S}<\epsilon_{1,A}<\epsilon_{2,S}
<\epsilon_{2,A}<\epsilon_{3,S}<\epsilon_{3,A}<...\ $. Without loss
of generality, we consider real eigenfunctions
$\phi_{\alpha,i}(x)$.

Due to the symmetry of the problem, it is useful to introduce a
complete set of orthonormal functions
\beq
\xi_{\alpha,R}(x) = {1\over \sqrt{2}} \big(
\phi_{\alpha,S}(x) + \phi_{\alpha,A}(x) \big) \; , \eeq 
and 
\beq \xi_{\alpha,L}(x) = {1\over \sqrt{2}} \big(
\phi_{\alpha,S}(x) - \phi_{\alpha,A}(x) \big) \; .
\eeq

If we fix the phase of $\phi_{\alpha,S}$ and $\phi_{\alpha,A}$, such that
$\phi_{\alpha,j}'(+\infty)<0$ for both $j=S$ and
$j=A$  {then $\xi_{\alpha,L}$ and $\xi_{\alpha,R}$ are
mainly localized in the left and right well respectively, at least for the
low lying states.

The field operator ${\hat \psi}(x)$ can be written as
\beq
\label{expansion} {\hat \psi}(x) = \sum_{\alpha=1}^{\infty}
\sum_{i=L,R} \xi_{\alpha,i}(x) \ {\hat c}_{\alpha,i}
\; ,
\eeq
in terms of the single-particle fermi operators ${\hat c}_{\alpha,i}$
(${\hat c}^{\dagger}_{\alpha,i}$)
satisfying the usual anticommutation rules, which destroy (create)
one fermion in the  state  $\xi_{\alpha,L}(x)$  or $\xi_{\alpha,R}(x)$.


By inserting Eq. (\ref{expansion}) into (\ref{ham}), the
Hamiltonian takes the form
\beq
{\hat H} = \sum_{\alpha=1}^{\infty} \sum_{i=L,R}
{\bar\epsilon}_{\alpha} \ {\hat c}_{\alpha,i}^{\dagger}
{\hat c}_{\alpha,i} -\sum_{\alpha=1}^{\infty}
J_{\alpha}({\hat c}^{\dagger}_{\alpha,L}{\hat c}_{\alpha,R}+
{\hat c}^{\dagger}_{\alpha,R}{\hat c}_{\alpha,L})
\label{hamiltonian}
\;, \eeq
where ${\hat n}_{F,\alpha,i}=
{\hat c}^{\dagger}_{\alpha,i}{\hat c}_{\alpha,i}$ is
the fermionic number operator of the $\alpha$-th state
in the $i$ well ($i=L,R$: $L$ means left and $R$ means right).
Notice that the diagonal part of
the Hamiltonian (\ref{hamiltonian}), given by
\beq
{\hat H}_0 = \sum_{\alpha=1}^{\infty} \sum_{i=L,R}
{\bar\epsilon}_{\alpha} \ {\hat c}_{\alpha,i}^{\dagger}
{\hat c}_{\alpha,i} \; ,
\label{h0}
\eeq
describes a double-well Fermi system with an infinite barrier,
i.e. with zero hopping terms. The energy
\beq
{\bar \epsilon}_{\alpha} =
{\epsilon_{\alpha,S}+\epsilon_{\alpha,A} \over 2} \;
\eeq
given by
\beq
\int \xi_{\alpha,i}(x) \left[-{\hbar^2\over 2m} {\partial^2\over
\partial x^2}
+ U(x)\right] \xi_{\alpha,i}(x) \ dx \; ,\eeq
{is the same for the left and right states described
by $\xi_{\alpha,L}(x)$ and $\xi_{\alpha,R}(x)$}
The energy \beq J_{\alpha}
={\epsilon_{\alpha,A}-\epsilon_{\alpha,S}\over 2} \eeq is the
energy of
hopping from the $L$ to the $R$ well within the
same doublet and gives directly half of the Rabi
frequency of the $\alpha$-th doublet as
\beq
\label{rabi}
\Omega_{\alpha}=\frac{J_{\alpha}}{\hbar} \; ,
\eeq
i.e. the Rabi angular frequency $\Omega_{\alpha}^{Rabi}$ is given by
$\Omega_{\alpha}^{Rabi} = 2\Omega_{\alpha}=2J_{\alpha}/\hbar$ \cite{sakurai}.

\section{Tunneling dynamics of free fermions in double well potential}

To study the dynamics sustained by the Hamiltonian
(\ref{hamiltonian}), we analyze the problem within the Heisenberg
picture. The Heisenberg equation of motion for the
operator ${\hat \psi}(x,t)$ is
\beq
i \hbar {\partial \over \partial t} {\hat \psi}(x,t) =
[{\hat \psi}(x,t),{\hat H}] =
\left[ -{\hbar^2\over 2m} {\partial^2\over \partial x^2}
+ U(x) \right] {\hat \psi}(x,t) \; ,
\eeq
and the time-dependent density operator is
\beq
{\hat n}_{F}(x,t) = {\hat \psi}^{\dagger}(x,t) {\hat \psi}(x,t) \; .
\eeq
The Heisenberg equations of motion for the operators
${\hat n}_{F,\alpha,L}$ and ${\hat n}_{F,\alpha,R}$ read
\beqa
\label{emnf1}
i\hbar\frac{d}{dt}{\hat n}_{F,\alpha,L}
&=& -J_{\alpha}({\hat h}_{\alpha,L}-{\hat h}_{\alpha,R}) \; ,
\\
\label{emnf2}
i\hbar \frac{d}{dt}{\hat n}_{F,\alpha,R}
&=& J_{\alpha}({\hat h}_{\alpha,L}-{\hat h}_{\alpha,R}) \; ,
\end{eqnarray}
where ${\hat h}_{\alpha,L}={\hat c}^{\dagger}_{\alpha,L}
{\hat c}_{\alpha,R}$, and
${\hat h}_{\alpha,R}= {\hat c}^{\dagger}_{\alpha,R}
{\hat c}_{\alpha,L}$. The equations of motion for the operators
${\hat h}_{\alpha,L}$ and ${\hat h}_{\alpha,R}$ are
\beqa
\label{emtf1}
i\hbar \frac{d}{dt}{\hat h}_{\alpha,L}
&=&-J_{\alpha}({\hat n}_{F,\alpha,L}-{\hat n}_{F,\alpha,R}) \; ,
\\
\label{emtf2}
i\hbar \frac{d}{dt}{\hat h}_{\alpha,R}
&=&J_{\alpha}({\hat n}_{F,\alpha,L}-{\hat n}_{F,\alpha,R}) \; .
\eeqa

We want to analyze the spatio-temporal evolution of our system
both at zero and finite temperature $T$. To this end, we evaluate the
thermal average of both sides of the Heisenberg equations of
motion obtained so far. The thermal average of an operator
${\hat A}(t)$ is given by
\beq
\langle {\hat A}(t) \rangle = {Tr\Big[
{\hat A}(t) {\hat \rho}_0  \Big] \over Tr\Big[ {\hat \rho}_0 \Big]
} = A(t) \; ,
\eeq
where ${\hat \rho}_0$ is the chosen statistical
density operator. It is clear that a thermal average evaluated by
using the density operator ${\hat \rho}=
e^{-\beta ({\hat H}-\mu {\hat N})}$ (with $\beta =1/(k_BT)$)
does not give rise to dynamics (time-dependent
observables) \cite{girardeau2}.
We want a statistical operator ${\hat \rho}_0$ which produces
the initial conditions
\beqa
n_{F,\alpha,L}(0) =
\langle {\hat n}_{F,\alpha,L}(0) \rangle =
f_{\alpha,L} \; ,
\label{initial1}
\\
n_{F,\alpha,R}(0) = \langle {\hat n}_{F,\alpha,R}(0) \rangle
= f_{\alpha,R} \; ,
\label{initial2}
\eeqa
with $0\leq f_{\alpha,L}\leq 1$ and $0\leq f_{\alpha,R}\leq 1$
parameters fixed by the choice of the statistical density operator.
We remark that $f_{\alpha,L}$ and $f_{\alpha,R}$ can be any
distribution. In particular they could be the initial
distributions corresponding to thermal equilibrium in the two
fully separate wells with different number of particles in each
well. In this case the statistical density operator is
\beq
\label{densita}
{\hat
\rho}_0 = e^{-\beta\,({\hat H}_0-\mu_{F,L} {\hat N}_{F,L} - \mu_{F,R} {\hat
N}_{F,R})} \ \; ,
\eeq
where ${\hat H}_0$ is given by Eq. (\ref{h0}),
${\hat N}_{F,i}=\sum_{\alpha=1}^{\infty}{\hat n}_{F,\alpha,i}$ and
$\mu_{F,i}$ the number operator and the chemical potential of fermions
in the $i$-th well, respectively. In this way we find
\beqa
f_{\alpha,L} =
{1\over e^{\beta\,({\bar \epsilon}_{\alpha} - \mu_{F,L})} + 1}
\; ,
\\
f_{\alpha,R} =
{1\over e^{\beta\,({\bar \epsilon}_{\alpha} - \mu_{F,R})} + 1}
\;.
\eeqa
The chemical potentials $\mu_{F,L}$ and $\mu_{F,R}$ are fixed by the number
of particles in the left and right wells at time zero:
\beqa
N_{F,L}(0) =
\sum_{\alpha=1}^{\infty}
{1\over e^{\beta({\bar \epsilon}_{\alpha} - \mu_{F,L})} + 1} \; ,
\\
N_{R,L}(0) =
\sum_{\alpha=1}^{\infty}
{1\over e^{\beta({\bar \epsilon}_{\alpha} - \mu_{F,R})} + 1} \;
\label{occupation} .
\eeqa

It is easy to show that the time-dependent density
profile $n_F(x,t)$ of the Fermi system,
\beq
n_F(x,t) = \langle {\hat \psi}^{\dagger}(x,t) {\hat \psi}(x,t) \rangle \; ,
\label{den}
\eeq
can be written as
\beq
n_F(x,t) = \sum_{\alpha=1}^{\infty} \sum_{i=L,R}
n_{F,\alpha,i}(t) \ \xi_{\alpha,i}(x)^2 \; ,
\label{den2}
\eeq
where $n_{F,\alpha,i}(t) =
\langle {\hat n}_{F,\alpha,i}(t) \rangle$.
In addition, from Eqs. (\ref{emnf1})-(\ref{emtf2}) it is
straightforward to get the following coupled ODEs
\beqa
&&{\ddot n}_{F,\alpha,L} + 2 \Omega_{\alpha}^2 \ n_{F,\alpha,L} =
2 \Omega_{\alpha}^2 \ n_{F,\alpha,R} \; ,
\label{ode1}
\\
&&{\ddot n}_{F,\alpha,R} + 2 \Omega_{\alpha}^2 \ n_{F,\alpha,R} =
2 \Omega_{\alpha}^2 \ n_{F,\alpha,L}
\label{ode2}
\eeqa
with $\Omega_{\alpha}$ given by Eq. (\ref{rabi}).

Notice that  the total number $n_{F,\alpha}= n_{F,\alpha,L}(t) +
n_{F,\alpha,R}(t)$ of fermions in the $\alpha$-th doublet  and
the fermionic hopping number $h_{\alpha}(t)=h_{\alpha,L}(t)+
h_{\alpha,R}(t)$ are both constants of motion.

It is not difficult to show that the ODEs (\ref{ode1}) and ({\ref{ode2})
with the initial conditions (\ref{initial1}) and (\ref{initial2})
have the following solutions
\beqa
n_{F,\alpha,L}(t) = f_{\alpha,L} \ \cos^2{(\Omega_{\alpha}t)} +
f_{\alpha,R} \ \sin^2{(\Omega_{\alpha}t)} \; ,
\\
n_{F,\alpha,R}(t) = f_{\alpha,R} \ \cos^2{(\Omega_{\alpha}t)} +
f_{\alpha,L} \ \sin^2{(\Omega_{\alpha}t)} \; .
\eeqa
The population imbalance $z_{\alpha}(t)$ within the $\alpha$-th double is
\beq
\label{zeta1}
z_{F,\alpha}(t) = n_{F,\alpha,L}(t)-n_{F,\alpha,R}(t) =
(f_{\alpha,L} - f_{\alpha,R}) \ \cos{(2\Omega_{\alpha}t)} \; ,
\eeq
and the total fermionic imbalance $Z_F(t)$ is given by
\beq
Z_F(t)={1\over N_F} \sum_{\alpha=1}^{\infty} z_{\alpha}(t) =
{1\over N} \sum_{\alpha=1}^{\infty}
(f_{\alpha,L} - f_{\alpha,R}) \ \cos{(2\Omega_{\alpha}t)} \; .
\label{zeta}
\eeq

\begin{figure}
\centerline{\includegraphics[width=9cm,clip]{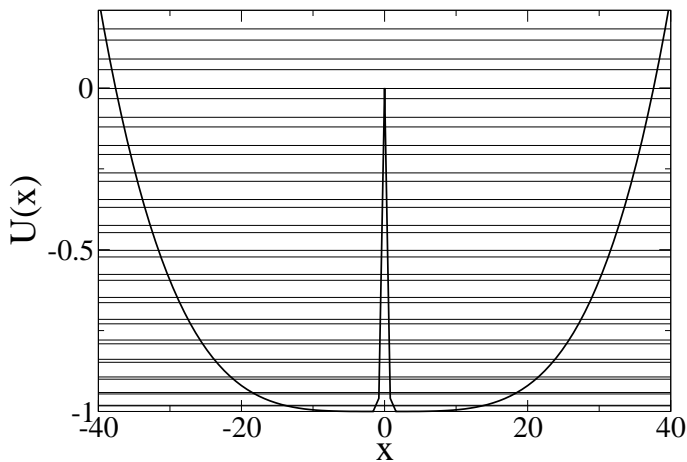}}
\caption{The double well potential (\ref{dw}) for $A=5 \cdot 10^{-7}$,
$B=1$,  and $C=5$.
For these values of the double-well potential parameters, there are $30$
energetic levels - corresponding to $15$ doublets - with energy smaller
than the height of the
barrier. Energies are measured in units of $\hbar \omega_{\bot}$, lengths in
units of $a_{\bot}=\displaystyle{\sqrt{\hbar/m\omega_{\bot}}}$.}
\label{fig1}
\end{figure}

We model the double-well trapping potential $U(x)$ (see Fig. 1) in the form
\beq
\label{dw}
U(x)=Ax^4+B(e^{-C x^2}-1)
\; . \eeq

\begin{figure}
\centering
\begin{tabular}{cc}
\epsfig{file=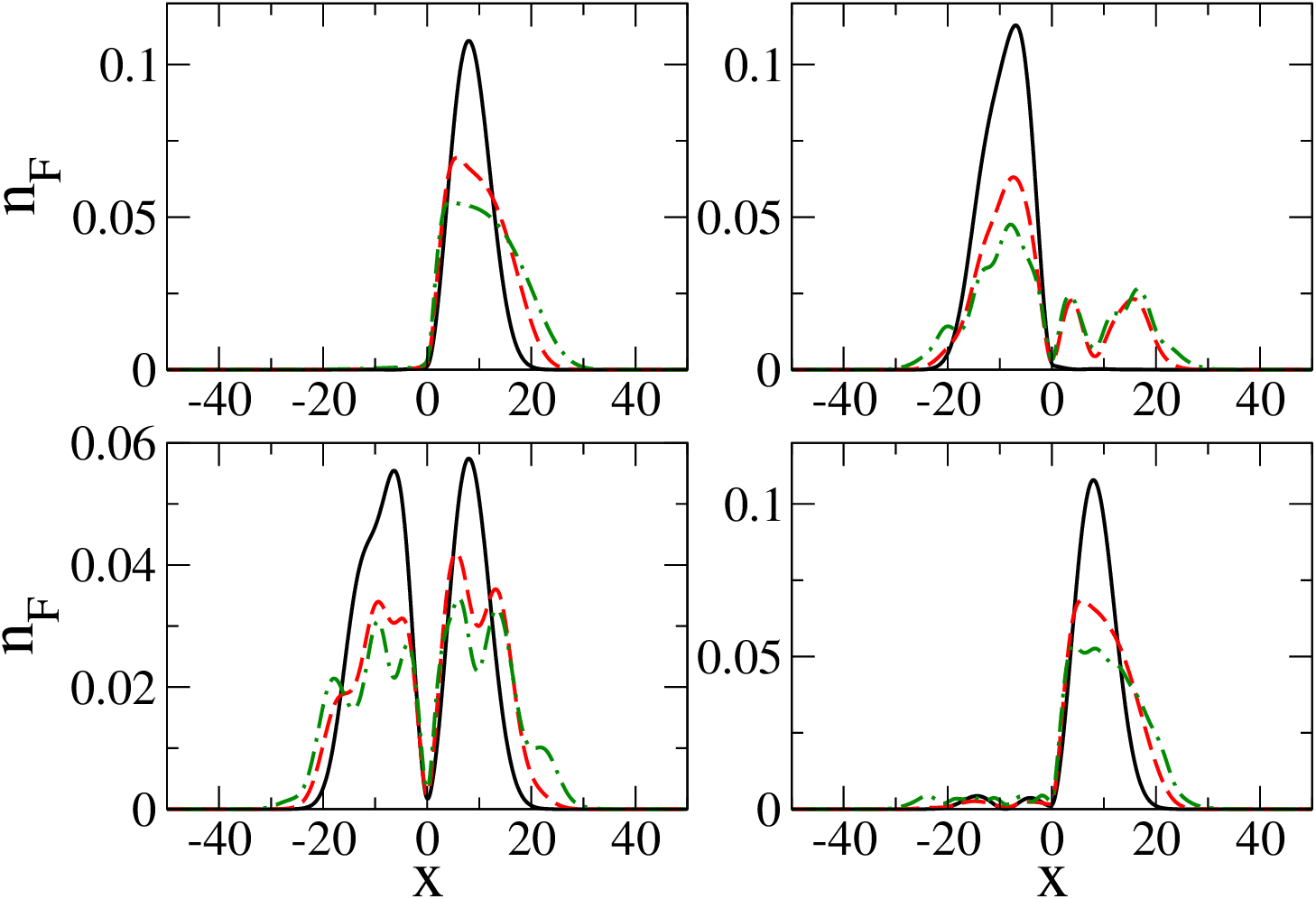,width=0.9\linewidth,clip=}\\
\epsfig{file=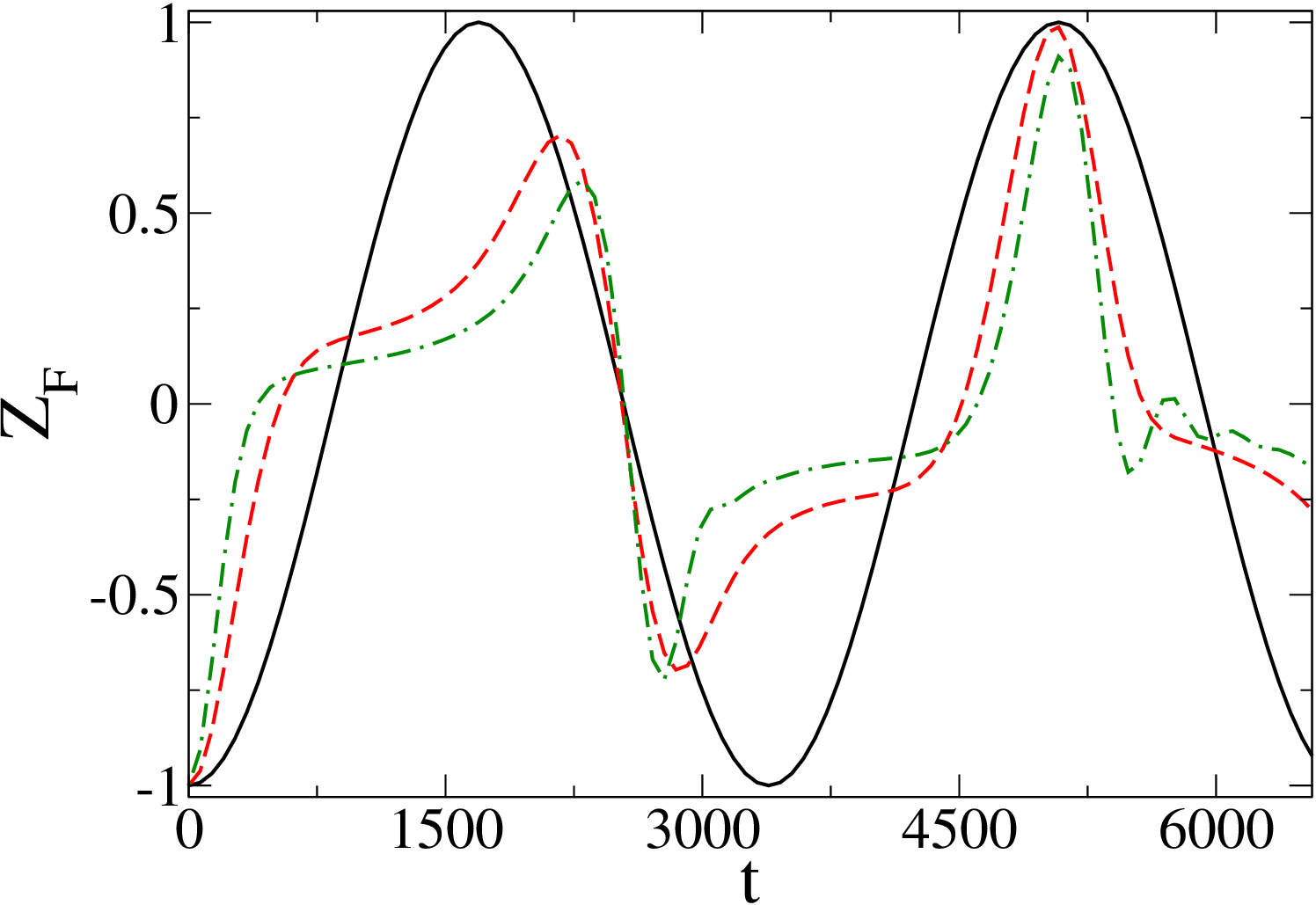,width=0.7\linewidth,clip=}
\end{tabular}
\caption{(Color online). Density profile $n_F$ vs.
space $x$ and total fermionic imbalance $Z_F$ vs. time $t$
with $N_F=1$. Plots of $n_F(x)$ are for different values of $t$.
We define $t_1$ as $2\pi/\Omega_1$. In
this figure: $t=0$ and $t=0.25\, t_1$ (the two upper panels of
$n_F$ from left to right); $t=0.88 \,t_1$ and $t=1.5
\,t_1$ (the two lower panels of $n_F$ from
left to right).
For both the $n_F(x)$ and $Z_F(t)$ plots,
the continuous line represents data for
$K_B \,T=0$, the dashed line
for $K_B \,T=0.038$ ($\simeq (\bar \epsilon_{2}-\bar \epsilon_{1})$), and
the dot-dashed line for $K_B T=0.087$
($\simeq(\bar \epsilon_{3}-\bar \epsilon_{1})$).
The plots at zero temperature are obtained with the fermion initially
localized in {the ground state of}
the right well; the
plots at finite temperature are obtained with $N_{F,L}(0)=0$ and $N_{F,R}(0)=
1$ {with level occupation defined by Eq. (\ref{occupation})}.
Times are measured in units of $(\omega_{\bot})^{-1}$, energies in units
of $\hbar \omega_{\bot}$, and lengths in units of $a_{\bot}=
\displaystyle{\sqrt{\hbar/m\omega_{\bot}}}$.
}\label{fig2}
\end{figure}

\begin{figure}
\centering
\begin{tabular}{cc}
\epsfig{file=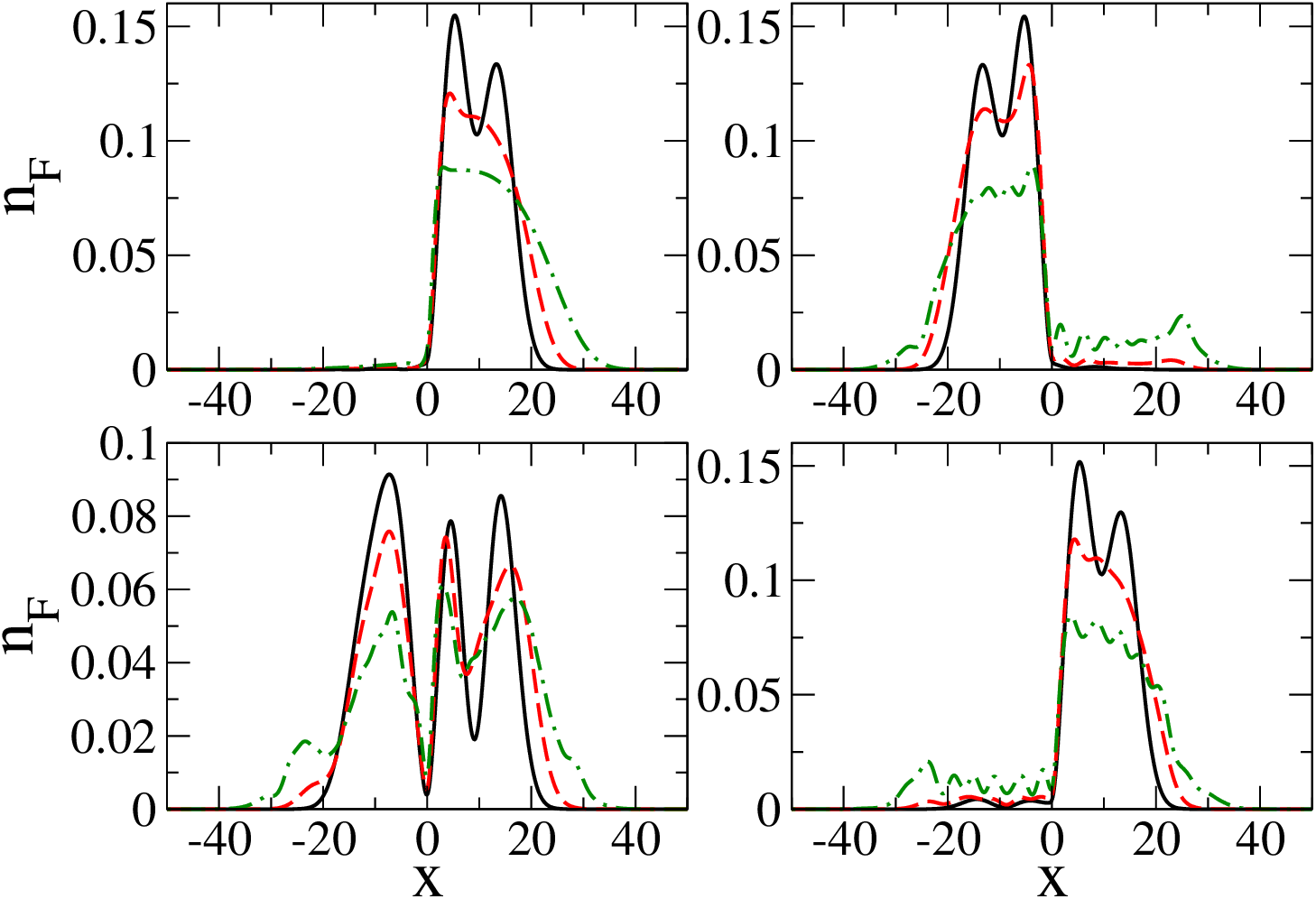,width=0.9\linewidth,clip=}\\
\epsfig{file=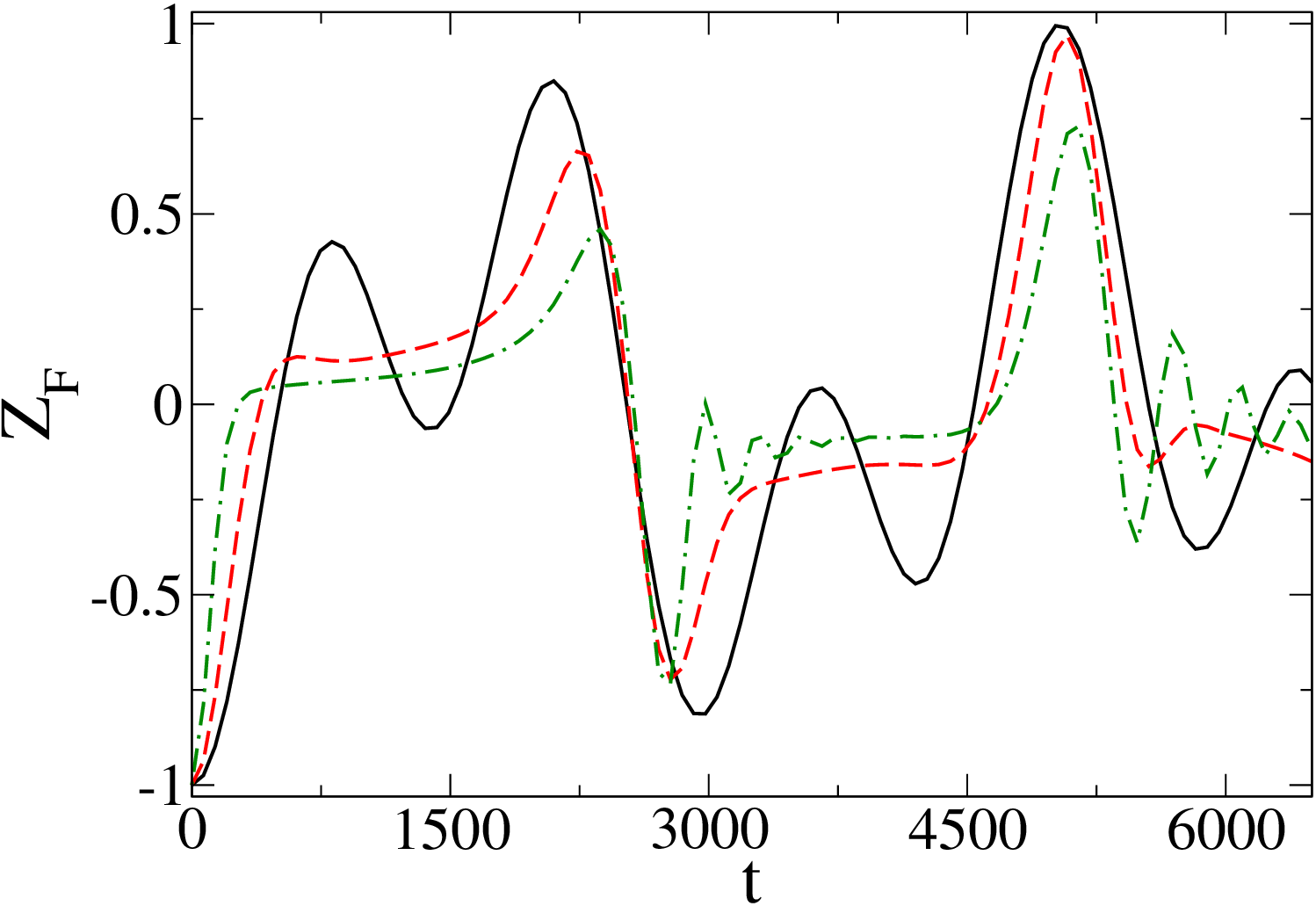,width=0.7\linewidth,clip=}
\end{tabular}
\caption{(Color online). Density profile $n_F$  vs.
space $x$ and total fermionic imbalance $Z_F$ vs. time $t$
with $N_F=2$. Plots of $n_F(x)$ are for different values
of $t$. We define $t_1$ as $2\pi/\Omega_1$. In
this figure: $t=0$ and $t=0.74\, t_1 $ (the two upper panels of
$n_F$ from left to right); $t=0.82 \,t_1$ and $t=1.5
\,t_1$ (the two lower panels of $n_F$ from left to right).
For both the $n_F(x)$ and $Z_F(t)$ plots, the continuous line represents data
for $K_B\,T=0$, the dashed line
for $K_B\,T=0.047$ ($\simeq(\bar \epsilon_{3}-\bar \epsilon_{2})$), and
the dot-dashed line for $K_B\,T=0.164$ ($\simeq(\bar \epsilon_{5}-\bar
\epsilon_{2})$).
The plots at zero temperature are obtained with
{the 2 fermions initially in the 2 lower states of}
 the right well; the
plots at finite temperature are obtained with $N_{F,L}(0)=0$ and
$N_{R,L}(0)=2$ {with level occupation defined by Eq.
(\ref{occupation})}. Units are the same as in Fig. 2.}
\label{fig3}
\end{figure}

\begin{figure}
\centering
\begin{tabular}{cc}
\epsfig{file=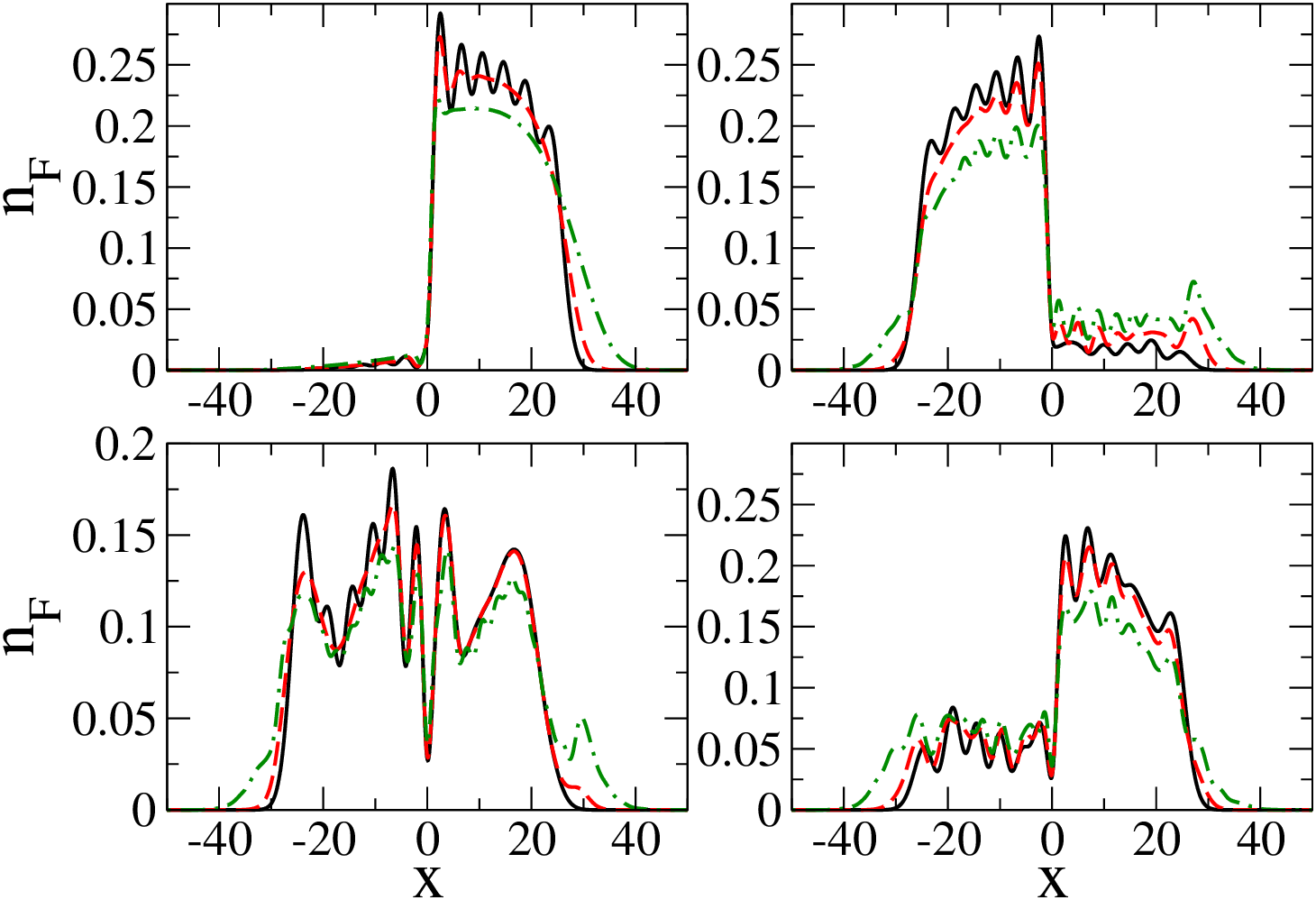,width=0.9\linewidth,clip=}\\
\epsfig{file=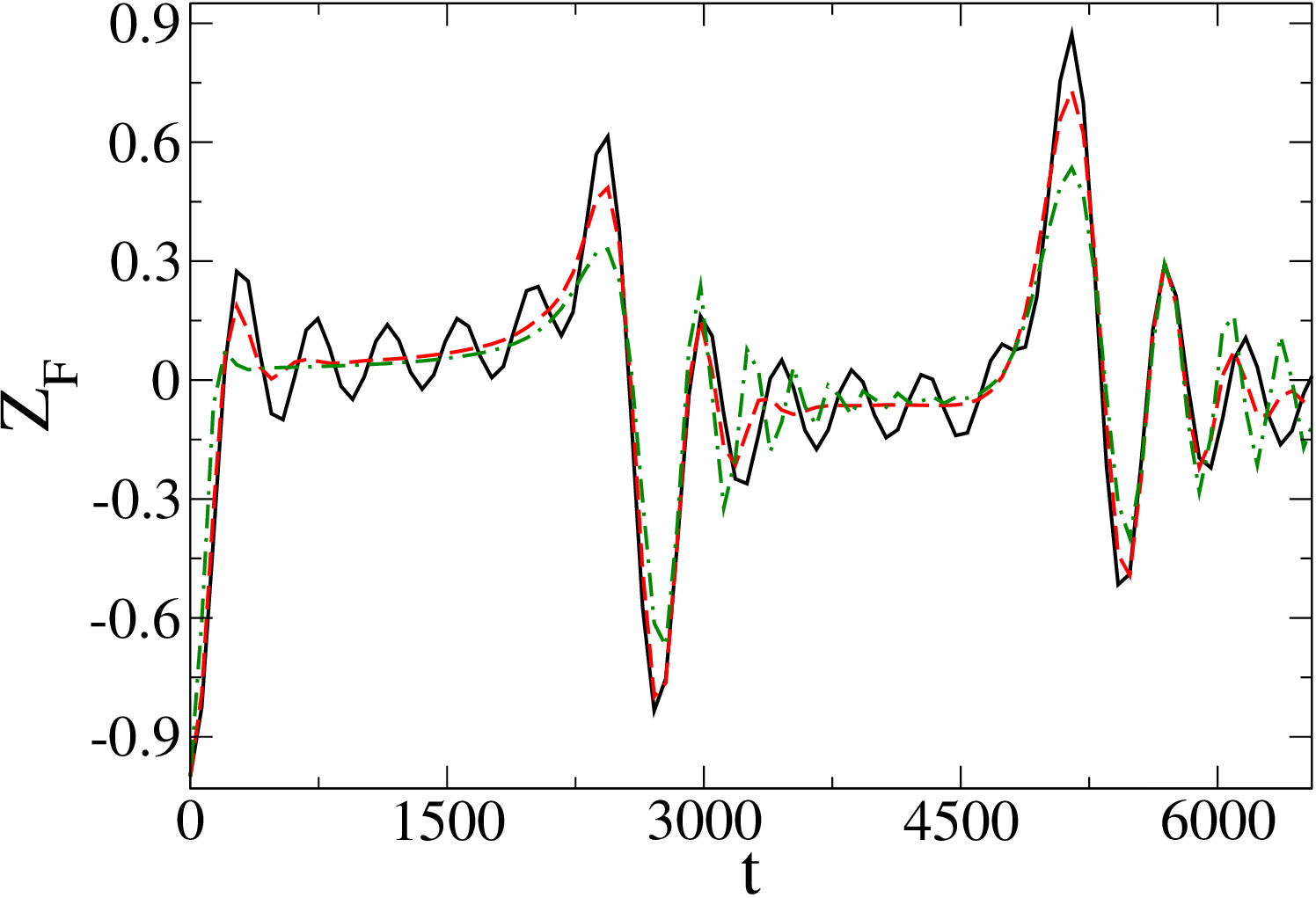,width=0.7\linewidth,clip=}
\end{tabular}
\caption{(Color online). Density profile $n_F$  vs.
space $x$ and total fermionic imbalance $Z_F$ vs. time $t$ with $N_F=6$.
Plots of $n_F(x)$ are for different values of $t$.
We define $t_1$ as $2\pi/\Omega_1$.
In this figure: $t=0$ and $t=0.76\, t_1$ (the two upper panels of
$n_F$ from left to right); $t=0.83 \,t_1$ and $t=1.52
\,t_1$ (the two lower panels of $n_F$ from left to right).
For both the $n_F(x)$ and $Z_F(t)$ plots, the continuous line represents data
for $K_B\,T=0$, the dashed line
for $K_B\,T=0.068$ ($\simeq(\bar \epsilon_{7}-\bar \epsilon_{6})$), and
the dot-dashed line for $K_B\,T=0.219$ ($\simeq(\bar \epsilon_{9}-\bar
\epsilon_{6})$).
The plots at zero temperature are obtained with
{the 6 fermions initially in the 6  lower states of}
 the right well; the plots at finite temperature are obtained
with $N_{F,L}(0)=0$ and $N_{R,L}(0)=6$ {with level occupation defined by Eq.
(\ref{occupation})}. Units are the same as in Fig. 2.}
\label{fig4}
\end{figure}

\begin{figure}
\centering
\begin{tabular}{cc}
\epsfig{file=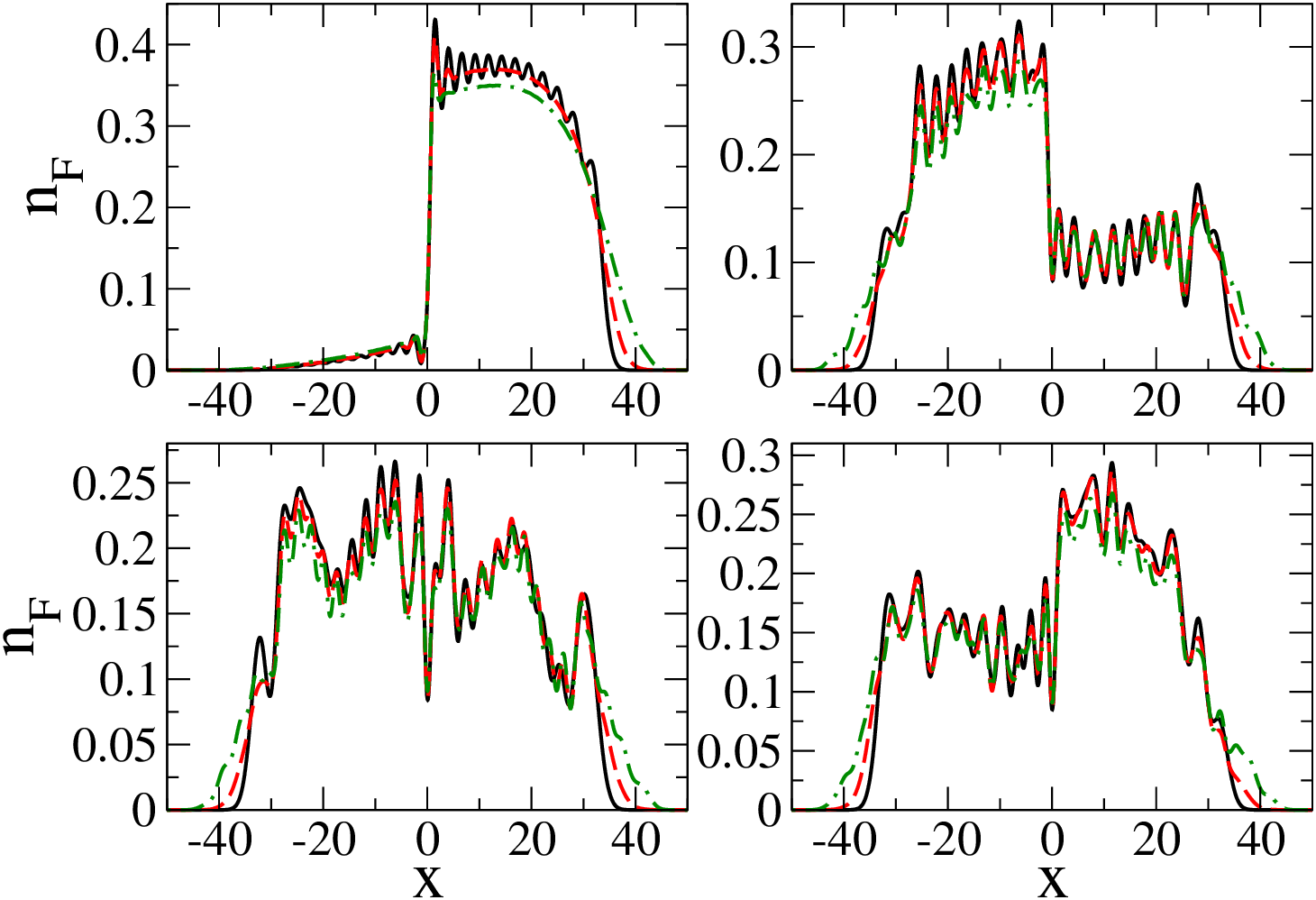,width=0.9\linewidth,clip=}\\
\epsfig{file=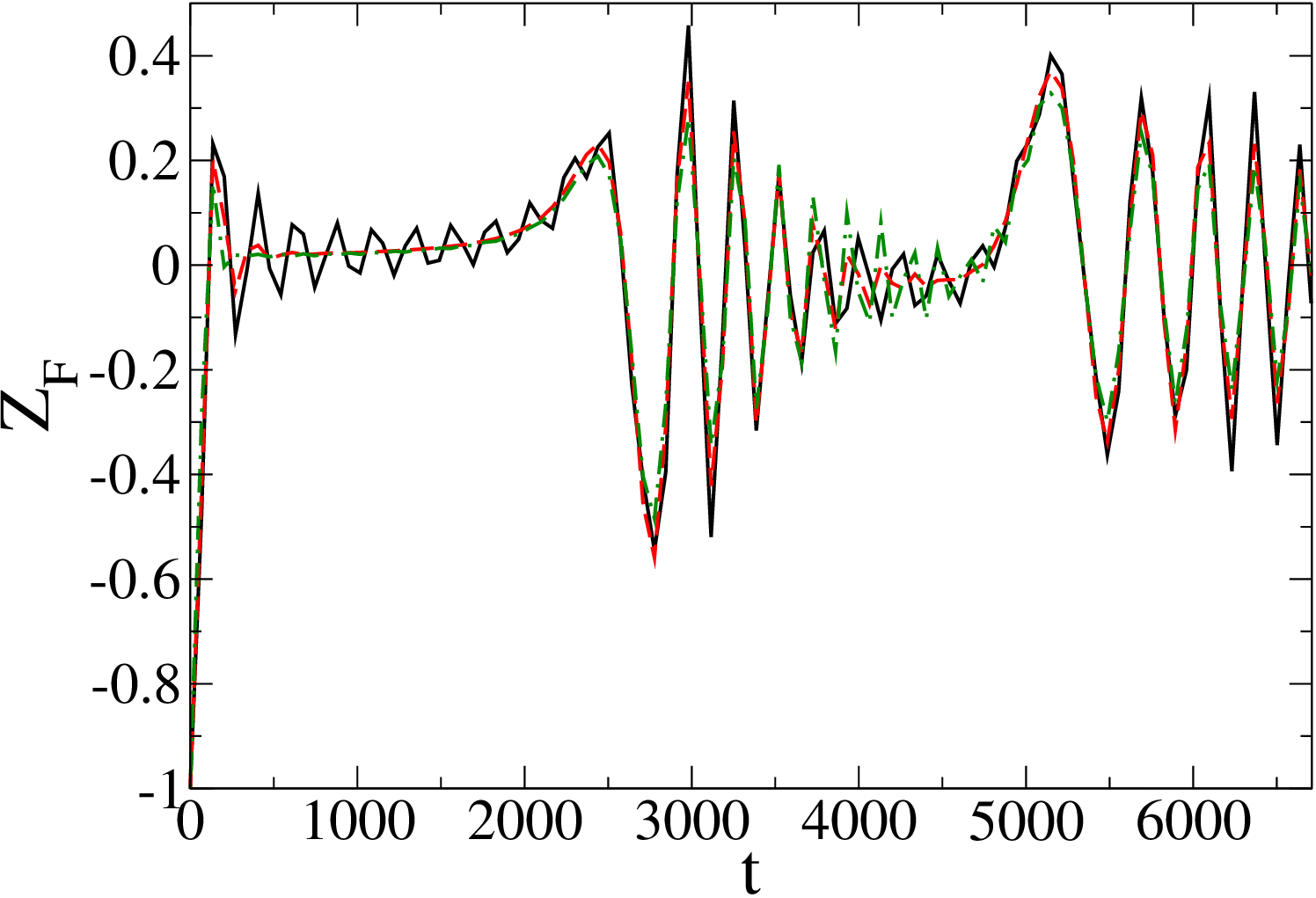,width=0.7\linewidth,clip=}
\end{tabular}
\caption{(Color online). Density profile $n_F$  vs.
space $x$ and total fermionic imbalance $Z_F$ vs. time $t$ with $N_F=12$.
Plots of $n_F(x)$ are for different values
of $t$. We define $t_1$ as $2\pi/\Omega_1$. In
this figure: $t=0$ and $t=0.77\, t_1$ (the two upper panels of
$n_F$ from left to right); $t=0.83
\,t_1$ and $t=1.52 \,t_1$ (the two lower panels of $n_F$
from left to right).
For both the $n_F(x)$ and $Z_F(t)$ plots, the continuous line
represents data for $K_B\,T=0$, the dashed line
for $K_B\,T=0.142$, ($(\bar \epsilon_{13}-\bar \epsilon_{12})<K_B\,T<(\bar
\epsilon_{14}-\bar \epsilon_{12})$), and
the dot-dashed line for $K_B\,T=0.333$, ($(\bar \epsilon_{15}-\bar
\epsilon_{12})<K_B\,T<(\bar
\epsilon_{16}-\bar \epsilon_{12})$).
The plots at zero temperature are obtained with
{the 12 fermions initially in the 12  lower states of}
 the right well; the
plots at finite temperature are obtained with $N_{F,L}(0)=0$ and
$N_{R,L}(0)=12$ {with level occupation defined by Eq.
(\ref{occupation})}. Units are the same as in Fig. 2.}\label{fig5}
\end{figure}

We study the time-dependent density profile (\ref{den}) and the total fermionic
imbalance (\ref{zeta}) both at zero and finite temperature for different number
of fermions. The results of this study are shown in Fig. 2 ($N_F=1$),
Fig. 3 ($N_F=2$), Fig. 4 ($N_F=6$), and Fig. 5 ($N_F=12$).

In our calculations we have supposed that the fermions of the system are
$^{40}$K atoms,
and set the harmonic transverse frequency $\omega_{\bot}$ to $160$ kHz, so that
$a_{\bot}\simeq 0.1 \mu$ m \cite{flavio2}.
The absolute temperatures used in Figs. 2-5 are calculated according to the above assumptions.
In each of these figures we show, for fixed values of $T$, the spatial
evolution of $n_F(x,t)$
at different times and the temporal evolution of $Z_F$. When $N_F=1$ and
$T=0$ - the
continuous line of Fig. 2 - only the lowest doublet is involved in the
dynamics. At zero
temperature, the temporal behavior of $Z_F$ is harmonic and is characterized
by only one frequency: $2 \Omega_1$. At finite temperatures - the dashed and
the dot-dashed lines of
Fig. 2 -  the fermions are allowed to
partially occupy doublets which are empty
at $T=0$ (see the captions of Fig. 2), and the harmonicity of
$Z_F$ is deformed by this thermal effect, as we can see from the $Z_F$ panel
of Fig. 2.

When the number of fermions is greater than one (Figs. 3-5), doublets other
than the lowest one
are involved in the dynamics. At $T=0$ the temporal evolution of the total
fractional imbalance  - the continuous line of the $Z_F$ plots - is no more
characterized by a single
frequency as in the $N_F=1$ case; it is, in fact, characterized by a mixing
of frequencies,
each of them proportional to the gap of
the corresponding
filled doublet. As in the $N_F=1$ case,
at finite temperatures
the fermions can partially occupy doublets higher in energy
than those filled at $T=0$ (see the captions of Figs. 3-5). The effect
 is to induce a deformation in the oscillations of the total
fermionic fractional imbalance - with respect to $T=0$ case - as
shown in the plots of $Z_F$ - the dashed and the dot-dashed lines - of
Figs. 3-5. The
thermal effects influence the fermionic density profile as well. In
particular, when the temperature is finite, $n_F(x)$ experiences a broadening
with respect to the case $T=0$.
From the plots of $n_F$ shown in Figs. 2-5, we see
that the higher is the temperature the wider is this broadening.

Finally, it is worth  observing that during the dynamics of our model
the system is not in thermal equilibrium and it will never reach it since
the particles do not interact between themselves. Each doublet
preserves its energy and its number of particles.

\section{Tunneling dynamics of spin-polarized fermions interacting with
intrinsically localized BEC}

In this section we consider the case of a spin-polarized quantum Fermi
gas in interaction  with a BEC which is intrinsically localized in one of
the two wells of the potential
(\ref{dw}) at zero-temperature.  In particular,  we are interested in the
changes in the Rabi frequencies
and in the fermionic imbalance dynamics induced by the presence of the BEC.
To this regard we restrict, for simplicity, to the case of a small number
of excited fermionic states present in the fermionic density
and to weak boson-fermion interactions. In this situation the condensate
will remain self-trapped (practically stationary) in the course of time
and  the fermionic imbalance dynamics will be qualitatively similar to
the one discussed in the previous section. The fermionic spectrum and the
 Rabi frequencies, however, will depend on the boson-fermion interaction
due to the presence of a bosonic effective potential in the fermionic
Schr\"{o}dinger equation of motion (see below).

To describe the spin-polarized fermionic gas  in interaction  with the BEC
we adopt a  mean field description  for the condensate but treat
the  fermions still quantum mechanically.  In this case one can show
\cite{karpiuk,ms05} that the dynamics of the mixture is described by the
following set of coupled equations
\beq
\label{mix}
i\hbar\frac{\partial\Psi}{\partial t} = \left[ 
-\frac{\hbar^2}{2m}\frac{\partial^2}
{\partial x^2} +U(x)+ g_B N_B |\Psi|^2+ g_{BF} n_F \right] \Psi \; , 
\eeq
\beq
\label{mix1}
i \hbar\frac{\partial \chi_j}{\partial t} = \left[ -\frac{\hbar^2}{2m}
\frac{\partial^2}{\partial x^2}
+ U(x) + N_B g_{BF} |\Psi|^2 \right] \chi_j \; ,
\eeq 
where $n_F(x,t) =\sum_{j=1}^{N_F}|\chi_j(x,t)|^2$
denotes the fermionic density
with $\chi_j(x,t)$ ($j=1,...,N_F)$ the set of orthonormal wave functions
which satisfy Eq. (\ref{mix1}),
$\Psi(x,t)$ is the bosonic wavefunction normalized to one and
such that $n_B(x,t)=N_B |\Psi(x,t)|^2$ is the bosonic density with
$N_B$ the total number of bosons. The 1D interaction strengths
are $g_B=2\hbar^2a_B/(ma_{\bot}^2)$ and $g_{BF}=2\hbar^2a_{BF}/(ma_{\bot}^2)$,
with $a_{B}$ and  $a_{BF}$ the
boson-boson and boson-fermion s-wave scattering lengths
and $a_{\bot}=\sqrt{\hbar/(m\omega_{\bot})}$
the characteristic length of the strong transverse harmonic
confinement of frequency $\omega_{\bot}$ \cite{flavio1,flavio2}.
A similar study for macroscopic quantum self-trapping Bose-Josephson
junction with frozen fermions
was done in \cite{kivshar09}. For simplicity, we also assumed equal masses
and the same trapping potential for both bosons and fermions. 
The bosonic cloud is self-trapped in one of the two wells 
when the nonlinear strength $g_B N_B$ exceeds a certain finite 
critical value \cite{smerzi}. This condition 
can be written as $|a_B|N_B/a_{\bot}>
(\epsilon_{1,A}-\epsilon_{1,S})/(\hbar \omega_{\bot})$ \cite{smerzi}, 
and it is fully satisfied in our numerical experiments. 

Note that the BEC wavefunction evolves according to a nonlinear
Gross-Pitaevskii equation (GPE) in which the fermionic density enters
as a potential, while the fermionic dynamics is still linear with the
condensate density playing the role of an additional external potential. 
From this it is clear that the fermionic eigenstates and eigenvalues can
be well approximated
by the linear Schr\"odinger equation
\begin{equation}
\label{ms}
i \hbar\frac{\partial \chi_j}{\partial t}= 
\left[ -\frac{\hbar^2}{2m}\frac{\partial^2}{\partial x^2} 
+ U_{eff}(x) \right] \chi_j \; ,  
\end{equation}
with the effective potential 
$U_{eff}(x) = U(x) + g_{BF} \bar n_B(x)$, with $\bar n_B(x)$ denoting the
stationary bosonic density. We have numerically verified that indeed 
the bosonic cloud is practically stationary. 
From Eq. (\ref{ms}) the possibility to
manipulate the fermionic dynamics by changing  $N_B$ or the inter-species
scattering length becomes evident.

\begin{figure}
\hskip 0.cm
\centerline{
\includegraphics[width=4.2cm,clip]{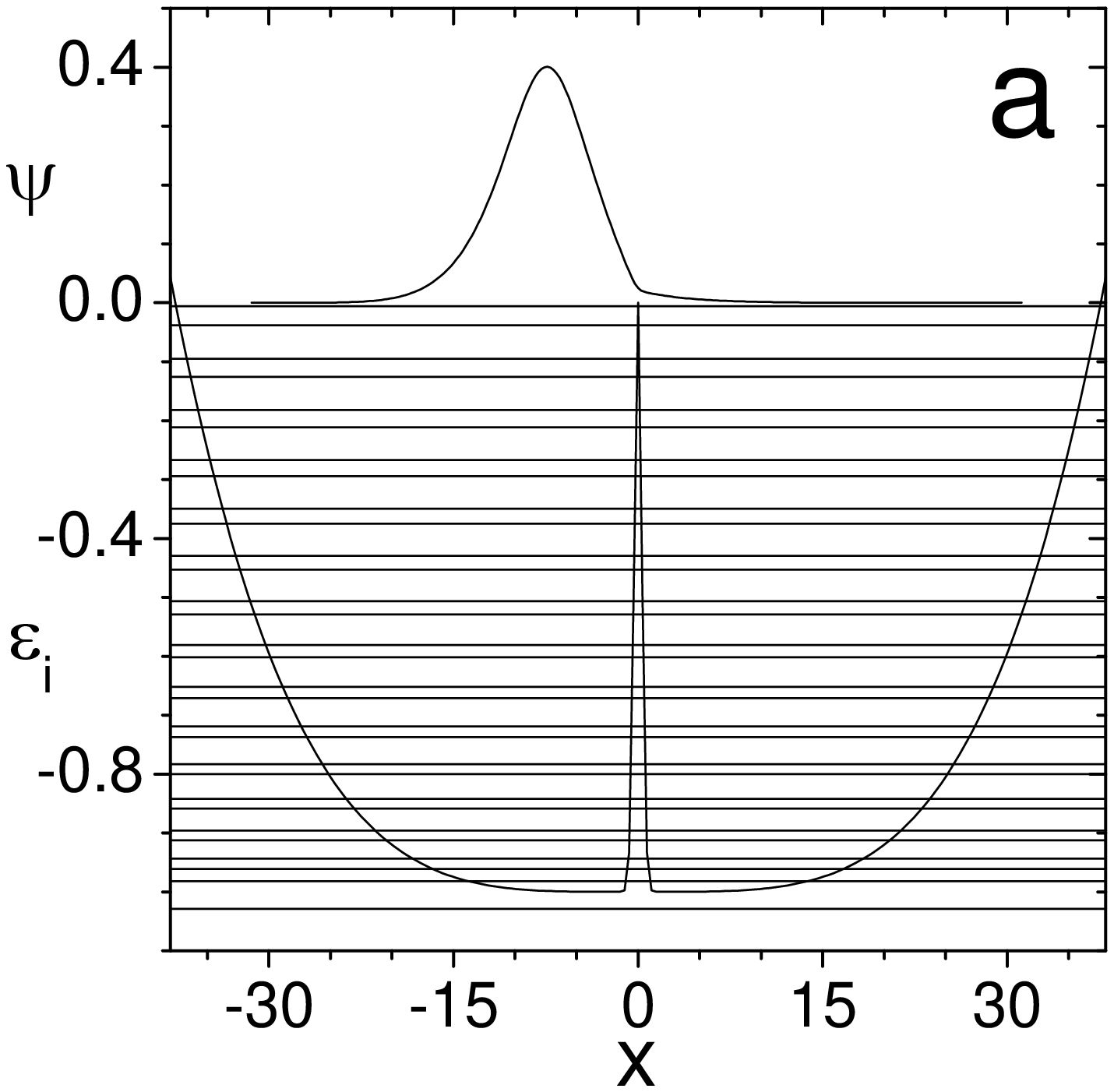}
\includegraphics[width=4.4cm,clip]{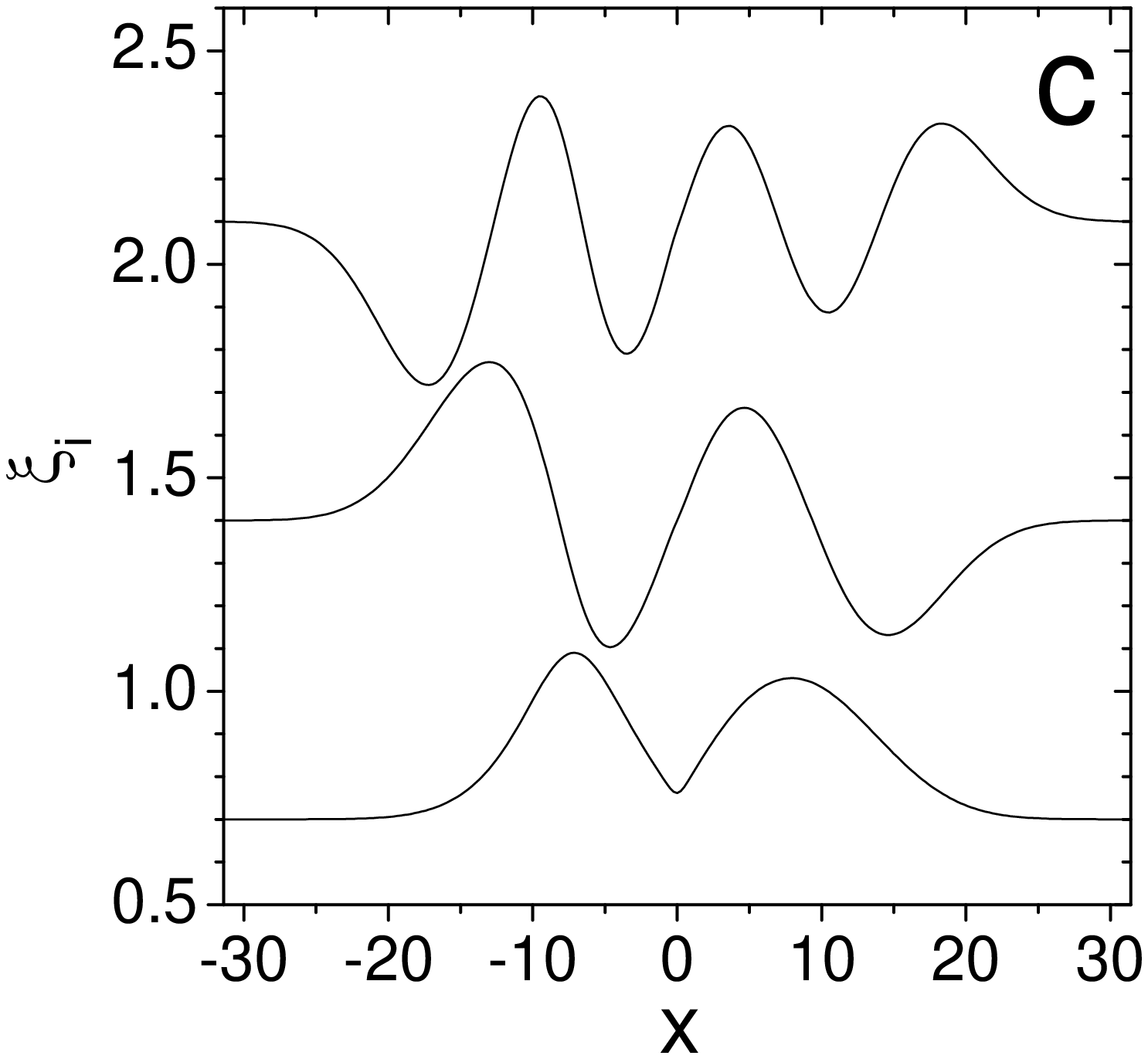}}
\centerline{
\includegraphics[width=4.3cm,clip]{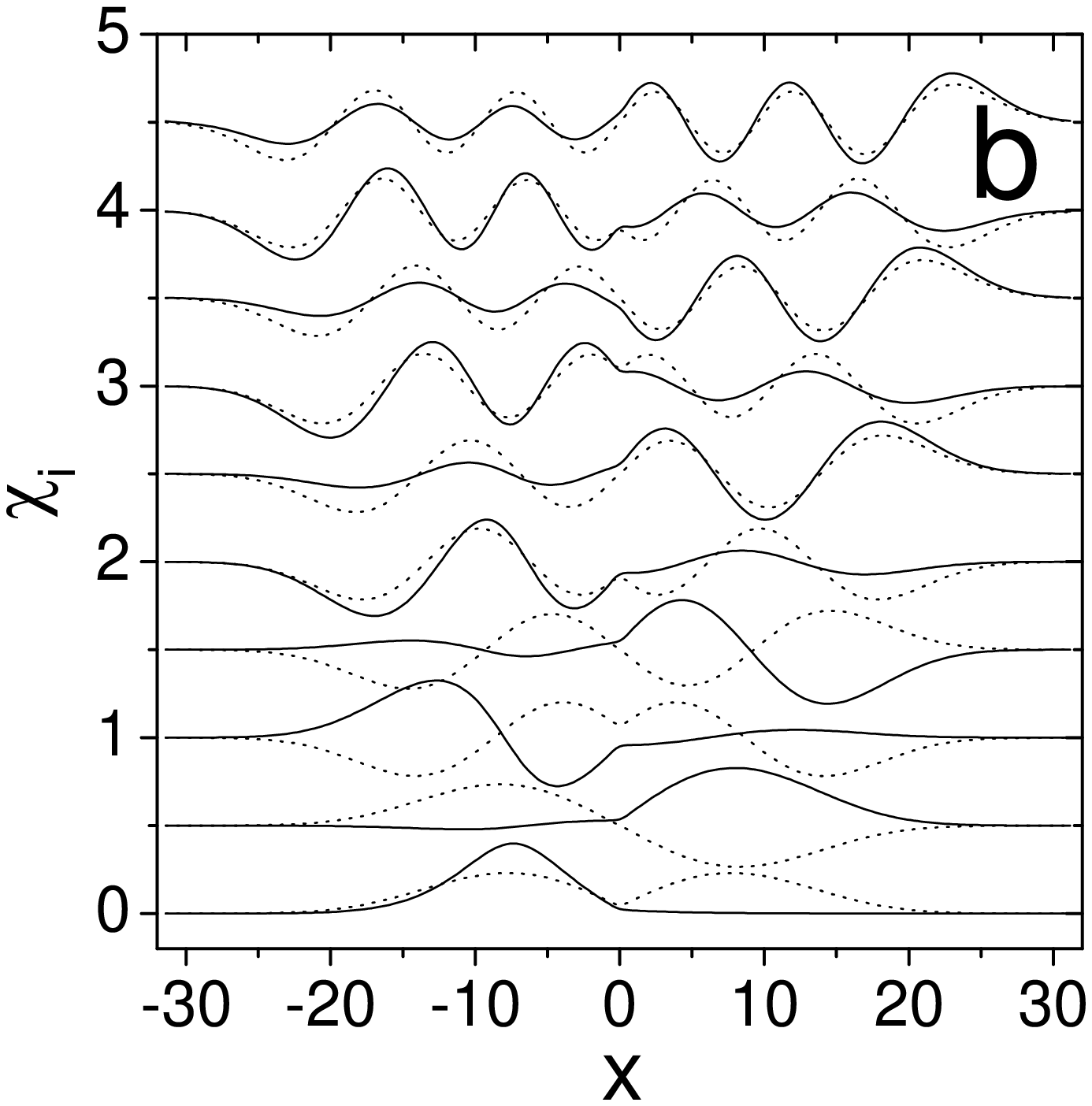}
\includegraphics[width=4.3cm,clip]{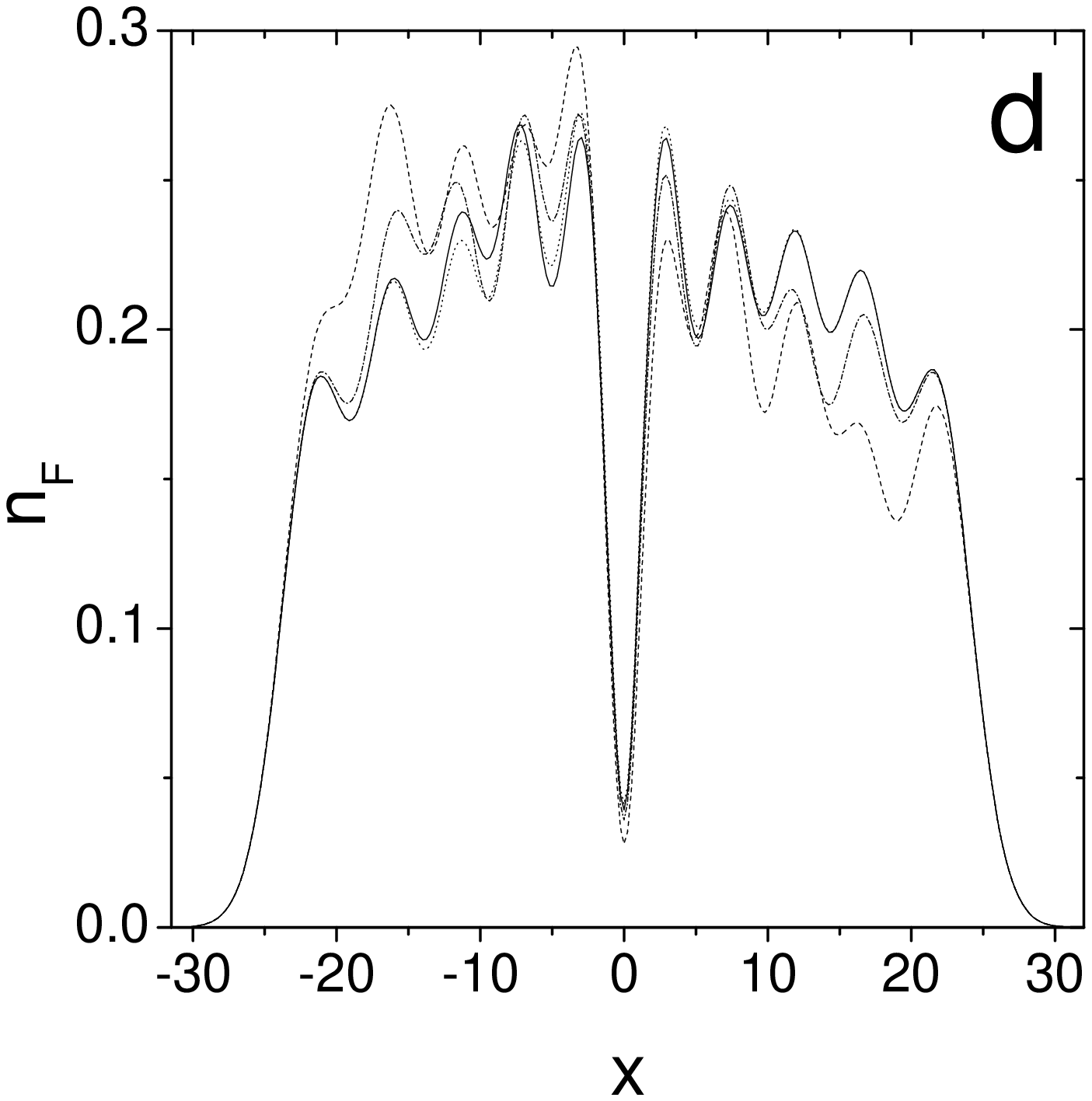}
}
\caption{
Panel a): condensate localized wavefunction $\Psi(x)$ (top curve)
corresponding to $N_B=470$ bosons in a  double well potential
and in the presence of $30$ spin polarized fermions with attractive
boson-boson and boson-fermion interactions $a_B=-0.001$, $a_{BF}=-0.001$.
Horizontal lines denote the first $30$  fermionic energy levels in the
presence of the condensate while the bottom curve represents the trapping
potential with parameters fixed as in Fig. \ref{fig1}.
Panel b): lowest ten fermionic eigenfunctions $\chi_i(x)$
(from bottom to top) in the presence (continuous curves) and
in the absence (dotted lines) of the BEC wavefunction depicted in panel
a) (an offset of $0.5$ between curves
has been added to avoid overlapping).
Panel c): first three excited states
in the presence of BEC constructed with the lowest eigenstates 
of panel b) (see text). 
An offset of $0.7$ between curves was added to avoid overlapping.
Panel d):  fermionic densities for 10 fermions including one (dotted line),
two (dash dotted line),
and three (dashed line) excited states depicted in panel c. The continuous
line denote the stationary fermionic density in absence of excited states.
Energies are measured in units of $\hbar \omega_{\bot}$, lengths in
units of $a_{\bot}=\displaystyle{\sqrt{\hbar/m\omega_{\bot}}}$.
The bosonic and fermionic wavefunctions are both normalized to one.}
\label{fig6ms}
\end{figure}

In order to check these predictions, we have numerically determined the
stationary states of the mixture  by solving in a self-consistent manner
the  time independent equations corresponding to  Eqs. (\ref{mix})
and (\ref{mix1}) \cite{ms05}  for the case of attractive interactions.
In Fig. \ref{fig6ms}a)
we show the stationary bosonic wavefunction  localized in the left well of
the potential and the first $30$ fermionic stationary levels obtained in
presence of BEC with the self-consistent method. Notice that the lowest
levels deviate from the corresponding ones obtained in Fig. \ref{fig1}
and the quasi degeneracy of the lowest levels is removed  due to the bosonic
effective potential (in the present case the Rabi frequencies of lowest
levels are increased due to the level splitting). In panel b)  of 
Fig. \ref{fig6ms} we show the first ten  fermionic stationary  wavefunctions 
corresponding to the lowest ten energy  levels of panel a), while in panel 
c) we depict the first three excited 
(e.g. non stationary) states, $\zeta_i(x)$, constructed from the lowest 
stationary wavefunctions as: $\zeta_i(x) 
= (\chi_{2i-1}(x)+\chi_{2i}(x))/\sqrt 2$. 
Note that the fermionic lowest energy eigenstate is asymmetric, 
having  the same localization of the condensate due to the attractive 
Bose-Fermi interaction, 
while the next level has the opposite localization. 
This implies that the lowest excited states are not localized 
in the same potential well, as for  the pure fermionic case considered before, 
but are extended between the two wells.

\begin{figure}
\hskip 1cm
\centerline{
\includegraphics[width=4.2cm,clip]{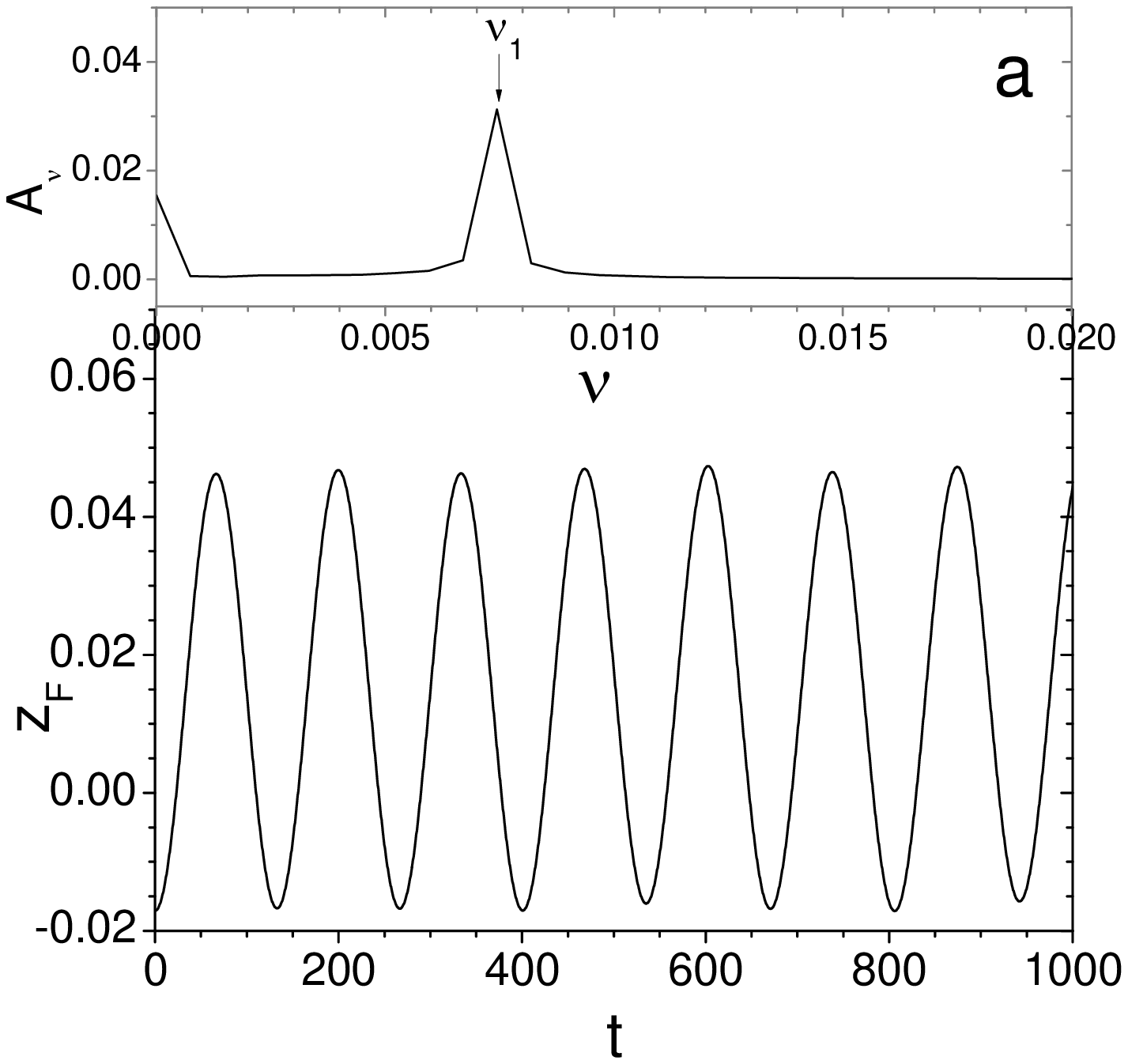}
\includegraphics[width=4.2cm,clip]{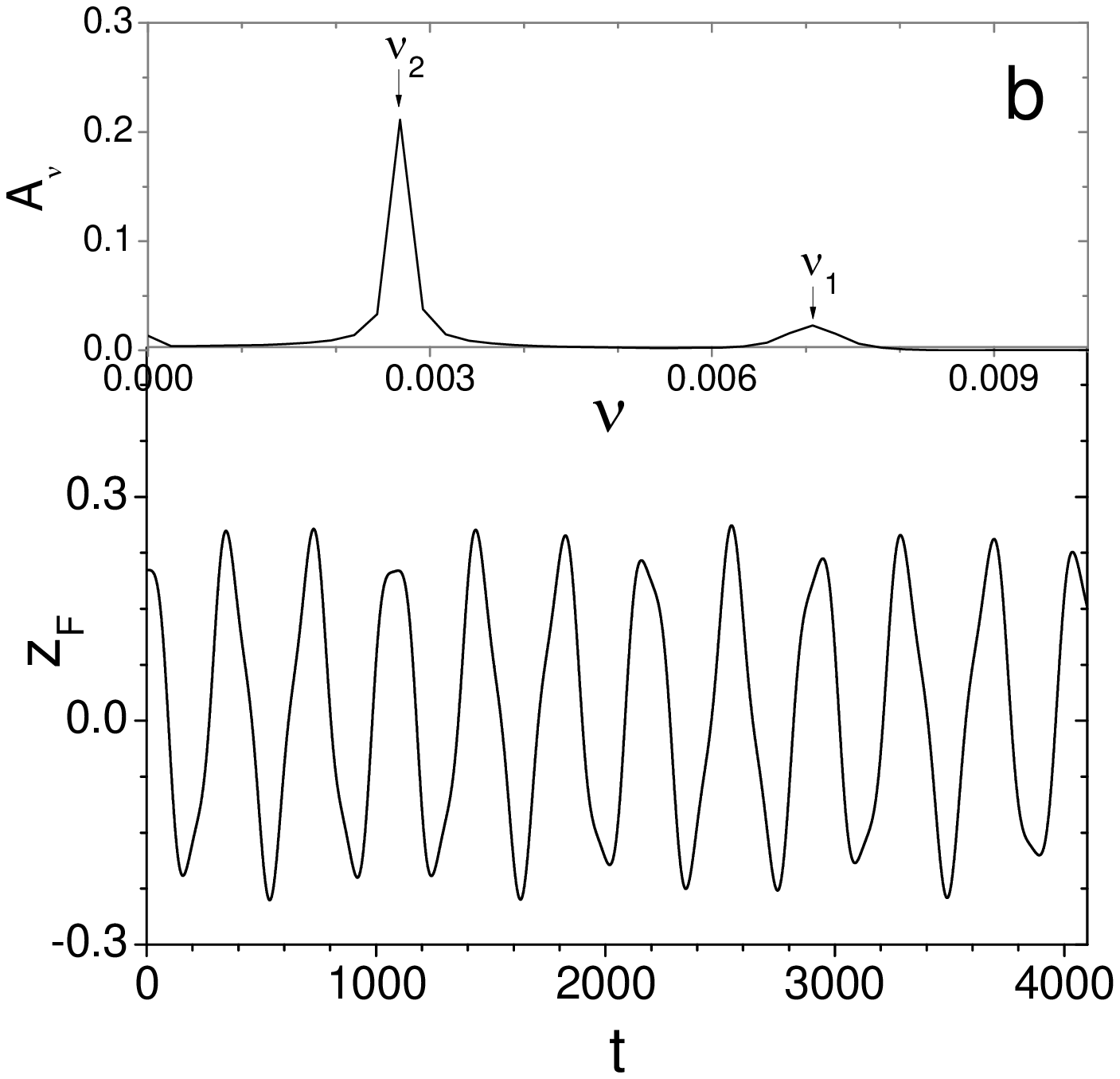}}
\centerline{
\includegraphics[width=4.2cm,clip]{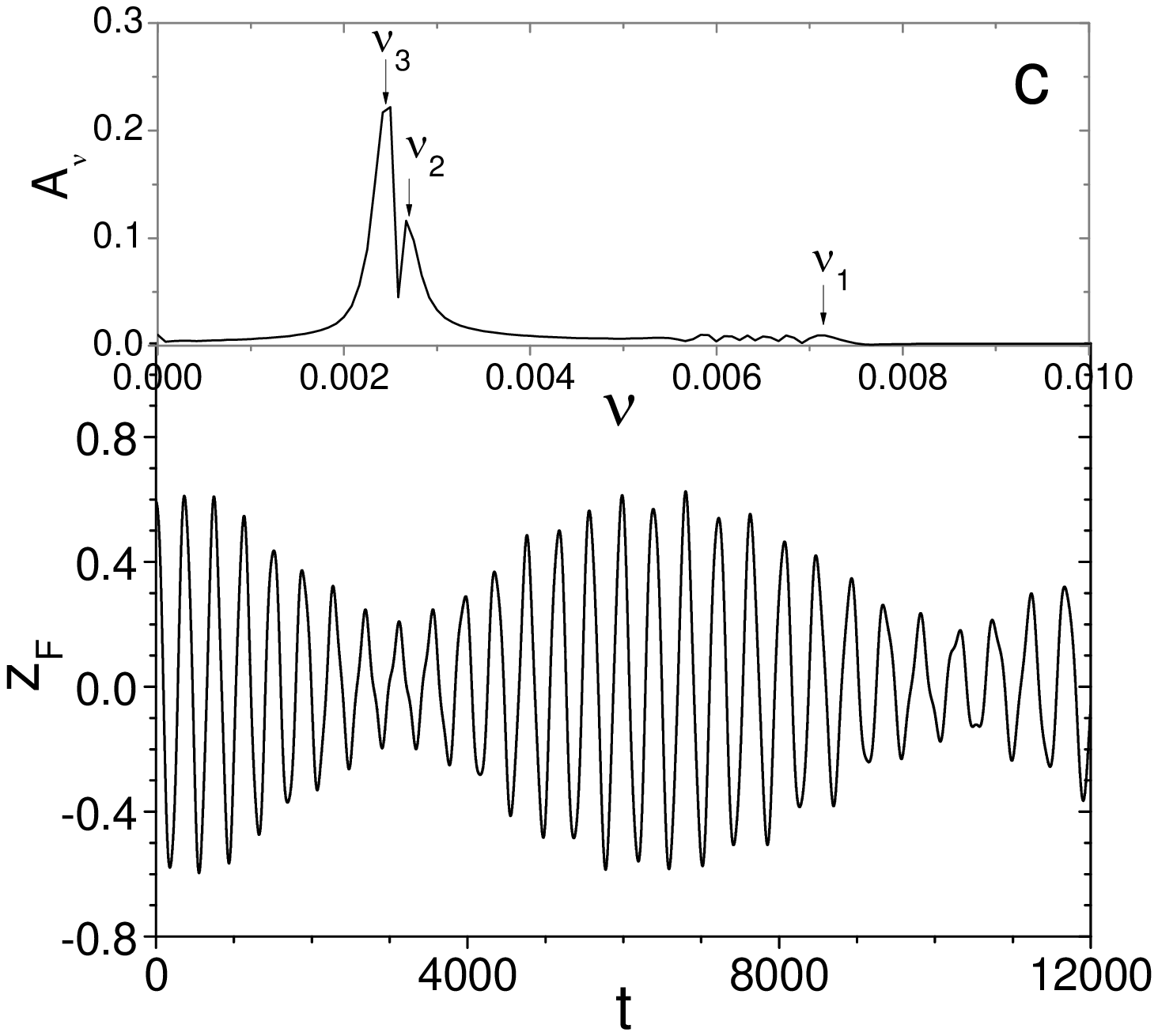}
\includegraphics[width=4.2cm,clip]{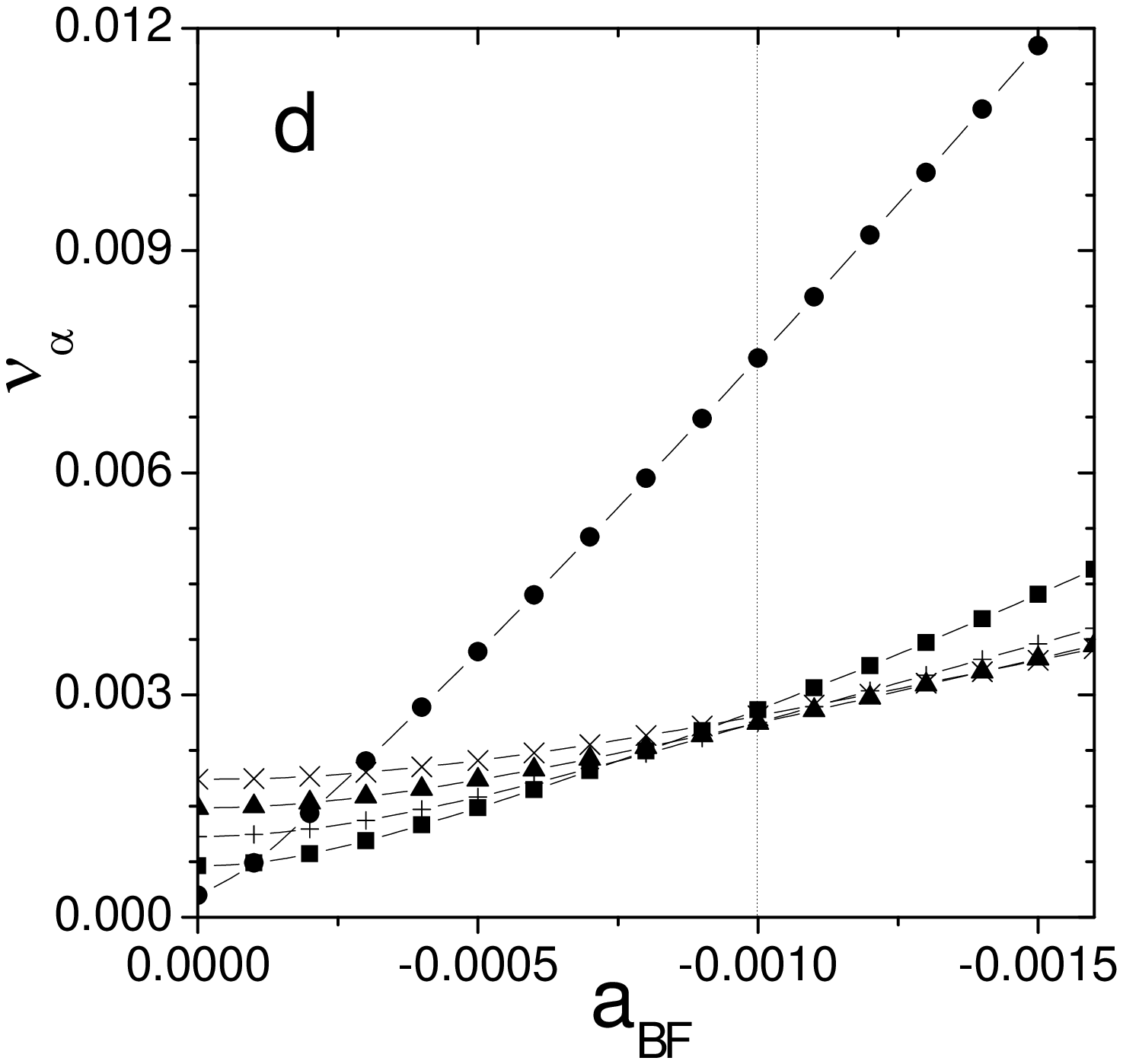}
}
\caption{
Panels a)-c): dynamics of the fermionic density imbalance $Z_F(t)$
(bottom) and corresponding Fourier spectrum (top part of panels)
as obtained from numerical integrations
of Eqs. (\ref{mix}) and (\ref{mix1}) using as
initial conditions the stationary localized bosonic wavefunction   and the
 non stationary fermionic densities
with one (a), two (b) and three (c) excited states  depicted in Fig.
\ref{fig6ms}.
Panel d): dependence of the Rabi linear frequency $\nu_{\alpha}$
on the boson-fermion interaction for the first
five excited levels indicated with dots, squares, plus, triangle and crosses,
respectively. The vertical line indicates the value of $a_{BF}$ at which the
other panels have been evaluated.
Parameters are fixed as in Fig. \ref{fig6ms}. Time is  measured in units of
${\omega_{\bot}}^{-1}$ and $a_{BF}$ is measured in units of $a_\bot$.}
\label{fig7ms}
\end{figure}

Excited fermionic densities corresponding to ten fermions and including
the lowest  excited states are shown in panel d) of Fig. \ref{fig6ms}
together with the stationary density. 
These excited densities were used, together with the BEC wavefunction
shown in the top part of panel 6a), as initial conditions to calculate
the total  fermionic imbalance density, $Z_F(t)$,  from direct numerical
integrations of Eqs. (\ref{mix}) and (\ref{mix1}).
The results are depicted in Fig.
\ref{fig7ms}, panels a)-c). From panel a) of this figure we see that when
only one excited state is present in the density, the imbalance dynamics
is periodic with  a single frequency in the spectrum. The 
Rabi linear frequency
$\nu_{\alpha}= \Omega_{\alpha}^{Rabi}/(2\pi)= \Omega_{\alpha}/\pi$
calculated from the dynamics of the fermionic imbalance
density is found in very good agreement (see panel d)) with the value
obtained from the energy spectrum and calculated from the  nonlinear
eigenvalue problem associated with Eqs. (\ref{mix}) and (\ref{mix1})
using the self-consistent approach \cite{ms05}. 
The same good agreement is 
found  in the case of two and three  excited states present 
in the  fermionic density panels (compare panels b),c), 
with corresponding values of panel d)). 

From panels 
a)-c) of Fig. \ref{fig7ms} it is clear that the fermionic imbalance dynamics
is given by  a superposition of harmonics in number equal to the number
of excited states and  with periods fixed  by the level spacing Rabi
frequencies. Also note from panel c) of Fig. \ref{fig7ms} the presence
of beatings of period $1/(\nu_2 - \nu_3)$ generated by  the the two close
frequencies $\nu_2$ and  $\nu_3$, as expected  for linear systems (notice
also the presence of small peaks close to the Rabi brequency $\nu_1$
probably due to the quasi-stationarity of the bosons).  

The dependence of
the Rabi frequency on the Bose-Fermi interaction is depicted in
Fig.\ref{fig7ms}d)  for the first five excited states. Note that the
frequency of the first excited state has a large 
variation with  $a_{BF}$ 
due to the fact that  the corresponding eigenstates are the ones most
effected by the presence of the condensate (see Fig. \ref{fig6ms}b).
Similar dependences are also obtained by changing the number of atoms
in the condensate and  keeping fixed the fixed boson-fermion interaction.

From this we conclude that in the BEC self-trapped  regime the dynamics 
for the fermionic density  imbalance remains qualitatively  similar to 
the one observed in the pure fermionic case but with the Rabi frequencies
explicitly dependent on the boson-fermion interaction. Since this interaction
can be easily changed by external magnetic fields via Feshbach resonances, 
we have found an effective way to  control the the dynamics of the spin 
polarized fermionic gas which can be implemented in a real experiment. 

\section{Conclusions}

We have investigated a confined dilute and ultracold spin-polarized gas
of fermionic atoms in a double-well potential. We have analyzed the quantum
tunneling through the central barrier by studying the density profile and the
total fermionic fractional imbalance. We have pointed out that despite the
fermions do not interact between themselves, the dynamics is quite 
complex, and it exhibits very interesting aperiodicity features. We have
performed our analysis both at zero and finite temperature by discussing
the role of the temperature in producing deformations of 
the density profile with respect to the zero-temperature case. 
We also have discussed the possibility 
to include in the system a bosonic component, which, under given hypothesis,
does not produce important changes in the dynamics of the system at
least within very low temperature regimes. In particular, we have shown
that the presence of a self trapped Bose-Einstein condensate
weakly interacting with the spin-polarized fermi gas allows to achieve
a quite effective control of the Rabi frequencies of excited fermionic
states by
changing the boson-fermion interaction with external magnetic fields via
Feshbach resonances.
This Bose-Einstein condensate-induced control of the fermionic quantum
dynamics
and can be implemented in a real experiment and can be of interest for
applications to quantum computing.
\\
This work has been partially supported by Fondazione CARIPARO
through the Project 2006: "Guided solitons in matter waves and
optical waves with normal and anomalous dispersion".
MS acknowledges partial support from MIUR through  a PRIN-2007
initiative.

\end{document}